\documentclass[12pt,a4paper,english,amssymb,nofootinbib,superscriptaddress]{revtex4-2}
\usepackage{lmodern}
\usepackage{lmodern}
\usepackage[T1]{fontenc}
\usepackage[latin9]{inputenc}
\usepackage{color}
\usepackage{babel}
\usepackage{verbatim}
\usepackage{amsmath}
\usepackage{amssymb}
\usepackage{esint}
\usepackage{graphicx}
\usepackage[style=english]{csquotes}
\usepackage[unicode=true,pdfusetitle,
 bookmarks=true,bookmarksnumbered=false,bookmarksopen=false,
 breaklinks=false,pdfborder={0 0 1},backref=false,colorlinks=true]
 {hyperref}
\hypersetup{
 linkcolor=blue,citecolor=blue}

\makeatletter

\pdfpageheight\paperheight
\pdfpagewidth\paperwidth

\@ifundefined{textcolor}{}
{%
 \definecolor{BLACK}{gray}{0}
 \definecolor{WHITE}{gray}{1}
 \definecolor{}{rgb}{1,0,0}
 \definecolor{GREEN}{rgb}{0,1,0}
 \definecolor{BLUE}{rgb}{0,0,1}
 \definecolor{CYAN}{cmyk}{1,0,0,0}
 \definecolor{MAGENTA}{cmyk}{0,1,0,0}
 \definecolor{YELLOW}{cmyk}{0,0,1,0}
}

\usepackage{babel}
\usepackage{babel}
\usepackage{babel}
\usepackage{babel}
\usepackage{babel}
\usepackage{babel}
\@ifundefined{definecolor}{color}{}
\usepackage{babel}

\usepackage{latexsym}\usepackage{bm}

\makeatother

\begin{document}
\title{Membrane paradigm approach to the Johannsen-Psaltis black hole
}

\author{Cagdas Ulus Agca}
\email{ulusagca@metu.edu.tr}

\author{Bayram Tekin}
\email{btekin@metu.edu.tr}

\affiliation{{\small{}Department of Physics,}\\
 {\small{}Middle East Technical University, 06800 Ankara, Turkey}}

\begin{abstract}
\noindent The Johannsen-Psaltis (JP) black hole is a phenomenologically viable metric obtained by judiciously deforming the Kerr black hole such that the metric is asymptotically flat and causal and is consistent with the weak field, post-Newtonian tests of gravity; however, it has additional hairs besides its mass and angular momentum. As it deviates from the Kerr black hole in the strong coupling regime, it is a useful metric to test the Kerr hypothesis that states that all astrophysical black holes are of the Kerr type. Here we give a membrane description of this black hole that effectively amounts to replacing the observable part of the black hole with a fluid with thermal properties. A timelike membrane, a stretched horizon local in time is assumed to exist. This membrane is expected to approximate the null event horizon that is highly non-local in time.  We derive the energy-momentum tensor of the fluid and all the transport coefficients using the action formulation to the membrane as advocated by Parikh and Wilczek. In the fluid description, one observes that the finiteness of the transport coefficients constrains the additional hairs of the Johannsen-Psaltis black hole. Analytically continuing the pressure of the fluid to all values of the radial coordinate $r$, one obtains interesting Van der Waals\textendash type behavior of the pressure of the fluid which diverges at the radius of the outer ergosphere, lending support, from the membrane paradigm's perspective to the claim that relativistic astrophysical jets are produced by the ergoregion of the black hole. 

\end{abstract}
\maketitle

\section{Introduction}

It is hard to make exact statements when macroscopic objects are concerned, yet we can make an exact statement about the astrophysical black holes (about all of them !) within the framework of general relativity: as long as they are isolated in an asymptotically flat universe, they are described by the Kerr metric \cite{Kerr}. This is sometimes called the {\it Kerr hypothesis} and of course it must be tested both in the weak field and strong field regimes. While the weak field regime has been tested in many different observations with no compelling evidence against the hypothesis, the strong field regime has only been recently brought into the domain of observation with gravitational waves emitted from binary black hole collisions \cite{LIGOScientific:2016aoc}, and the Event Horizon Telescope that captured the pictures of environments of supermassive black holes  \cite{EHT1, EHT2}. One rather remarkable property of the Kerr metric is the following: all of its multiple moments are related to the lowest order two moments: its mass ($m$) and spin ($J$). This fact is best explained by the Geroch-Hansen multipole moment expression \cite{Geroch1, Hansen1} in the units $G=c=1$: 
${\mathcal{M}}_\ell:= m_\ell+  i s_\ell= m( ia)^\ell$, where $m_0 = m$ and $s_1= m a = J$. This relation, sometimes dubbed as the "no-hair theorem" \cite{no_hair_theorem} will not be valid even for the slightest deviation of the metric or for any other compact object. So testing the Kerr hypothesis would in the end boil down to testing the no-hair theorem.

Most of the deformations of the Kerr metric yield pathological results such as the existence of closed timelike curves, or naked singularities. Moreover, because of the powerful uniqueness theorems of the Kerr black hole \cite{Uniqunesstheorem} within general relativity, it is just in vain to modify the Kerr metric as a solution to Einstein's equations: one must either consider alternative theories of gravity that allow viable deformations of the Kerr metric in the strong field regime or, perhaps better yet, without referring to any particular field equations, consider phenomenological metrics that judiciously deform metric.  In the literature, one can find many works devoted to the first type of deformations, that is modified Kerr solutions (albeit almost always approximate solutions) to modified theories of Kerr gravity, and not so many phenomenological metrics except the ones constructed by Johannsen and Psaltis {\cite{Johannsen_2013,A_metric_testing_johannsen}}. Here we shall study these metrics in the context of the membrane paradigm approach to black holes which is an effective description of the black hole in terms of a fluid with various thermal properties. This is an exotic fluid (which may not be easily drinkable as it has a negative bulk viscosity) that mimics the true event horizon of the black hole.  Our motivation to study the modified Kerr metric in the membrane paradigm was to understand if there are some constraints on the deformations parameters, and indeed there are as we shall see. In some sense, the theory is still a great 
 \enquote{experimental tool} and a viable membrane paradigm that was quite useful in understanding various aspects of black holes in general relativity, helping us constrain the deformation parameter, indeed a new hair, in the parametrically deformed Kerr metric.

The layout of the paper is as follows: In Sec. II we give a detailed account of the membrane construction using the action formalism of Parikh-Wilczek \cite{Parikh_original}; Sec. III is the bulk of our paper: we construct the membranes for the JP metric after considering the membrane for its static limit. 
 The computations presented here can be applied verbatim to any given stationary metric that forms an asymptotically flat deformation of the Kerr black hole.

\section{Construction of a Gravitational Membrane}
Our main goal is to construct a gravitational membrane that effectively reproduces the observable properties of the JP metric. Since this metric is somewhat cumbersome, it pays to first revisit the membrane paradigm in the easier context of the modified Schwarzschild metric (that is the zero rotation limit of the JP metric) to introduce the necessary physical ingredients. We claim no novelty in the basic construction of the membrane, the following is a recapitulation of this already-established tool.

\subsection{Basics of the membrane paradigm}
The {\it event horizon } of a static or stationary black hole is a null hypersurface of codimension one with a degenerate metric. It is highly nonlocal in time and hence not observable to transient observers like us and is not a tangible concept for astrophysical purposes. To remedy this and to approximate the black hole as a local object in time, an effective description that makes use of timelike surfaces, not null surfaces is needed. For this purpose, 
a \enquote{stretched horizon} or \enquote{fluid bubble,} that is, a timelike surface \cite{10_Action} with electrical conductivity having a resistance of 377 $\Omega$, shear and bulk viscosities \cite{Damour_Actio_membranes} arbitrarily close to the event horizon was introduced by Damour \cite{Damour_Actio_membranes}. This pragmatic point of view, that is endowing a black hole with a timelike facade has proven to be very useful. A careful calculation on the horizon led to Ohm's law \cite{TheMembraneModelofBlackHolesandApplications}, Joule's law, and the nonrelativistic Navier-Stokes equation \cite{MembraneHorizonsBlackHolesNewClothes}. Early works of Damour, followed by Thorne and Price \cite{membrane_THORNE} showed a way to mimic the null horizon in a well-defined approximate way since the membrane is not null by nature, it has a nondegenerate Lorentzian metric which allows computations local in time. This approach to black hole dynamics was coined as the membrane paradigm of black holes \cite{membrane_THORNE}. More recently \cite{Parikh_original}, a proper action formulation of the membrane was given that makes the computations rather straightforward, and hence we shall use this. See \cite{agca_2023} for a recent review of these ideas.

\subsection{Quantities describing the fluid membrane}

Let $\mathcal{H}$ be the event horizon of a stationary black hole, then there exists, by definition, a null geodesic vector field $\ell$ that generates $\mathcal{H}$.  One can define a timelike stretched horizon $\mathcal{H}_s$ arbitrarily close to this event horizon. In the Arnowitt\textendash{}Deser\textendash{}Misner (ADM) decomposition of the metric \cite{ADMoriginal}, let 
$N$ be the lapse function that can be chosen such that in the limit $N\rightarrow 0$, the stretched horizon goes to the true horizon, that is $\mathcal{H}_s|_{N \rightarrow 0}=\mathcal{H}$ \cite{MembraneHorizonsBlackHolesNewClothes}. This limit in the geometry cannot be smooth and hence $N$ will be used as a regulator in various geometric objects as we shall see below.

Let $(\mathcal{M},g)$ be the $(3+1)$-dimensional total black hole spacetime and $(\mathcal{H}_s,h)$ be a $(2+1)$-dimensional submanifold of $(\mathcal{M},g)$ with the pullback metric $h=\phi^* g$, which can also be considered as the projector $h^\mu\,_\nu:T_p(\mathcal{M})\longrightarrow T_p(\mathcal{H}_s)$. A spacelike cross section of $\mathcal{H}_s$ is also a submanifold of dimension two, which we denote as  $(\Sigma,\gamma)$. We can do differential geometry adapted to these two submanifolds, which boils down to a $(2+1+1)$ splitting of the full spacetime. Let $\nabla_\mu$ be the $g$-compatible covariant derivative and $D_\mu$ be the $h$-compatible covariant derivative while $\mathcal{D}_\mu$ be the $\gamma$-compatible covariant derivative \cite{Parikh_original}. Then, let $V^\mu\in T_p(\mathcal{M})$ and $n^\mu$ be a spacelike unit vector normal to $\mathcal{H}_s$, then defining the extrinsic curvature of $\mathcal{H}_s$ as $K_{\sigma\mu} :={h}{^\sigma}_{\mu}\nabla_\sigma n^\mu$, one has the identity   
\begin{equation}
    {h}{^\sigma}_{\mu} \nabla_\sigma V^\nu= D_\mu V^\nu-K_{\sigma\mu} V^\sigma n^\nu.
    \label{extrinsiccurvaturedefinition}
\end{equation}
 Let $u^\mu$ be the unit normal timelike vector to $\mathcal{H}_s$ chosen to satisfy $u^\mu n_\mu=0$. $u^\mu$ can be considered to be the four velocity of a fiducial observer with proper time $\tau$.  
Here is the crucial part of the discussion: for the membrane paradigm to approximate the true horizon, a nullness constraint on the stretched horizon should be imposed; this means a change of character of $u^\mu$ and $n^\mu$ in the limits
\begin{equation}
    N\, u^\mu\longrightarrow \ell^\mu, \hskip 2 cm N\, n^\mu\longrightarrow \ell^\mu,
\end{equation}
 as $N\rightarrow 0$, $\ell^\mu \ell_\mu=0$.

The relations we shall use for the membrane paradigm can be summarized as \cite{Parikh_original}
\begin{eqnarray}
    &&\ell^\mu \ell_\mu=0,\quad\ n^\mu n_\mu=1,\quad
  u^\mu u_\mu=-1,\quad u^\mu n_\mu=0,\quad
  a^\mu=n^\gamma\nabla_\gamma n^\mu=0,\nonumber \\
 && K_{\mu\nu}n^\nu=0,\quad
   {K}{^\gamma}_{\mu}= {h}{^\gamma}_{\nu}\nabla_\gamma n^\mu, \quad
   \lim_{N\to 0} N\, u^\mu=\ell^\mu, \quad\lim_{N\to 0} N\, n^\mu=\ell^\mu.\notag\\
   &&{h}{^\mu}_{\nu}={\delta}{^\mu}_{\nu}- n^\mu n_\nu,\quad {\gamma}{^\mu}_{\nu}={h}{^\mu}_{\nu} + u^\mu u_\nu={\delta}{^\mu}_{\nu}-n^\mu n_\nu+u^\mu u_\nu,\quad u^\mu=\left(\frac{d}{d\tau}\right)^\mu.
\end{eqnarray}
\subsection{An action formalism for the membrane paradigm}

Parikh and Wilczek \cite{Parikh_original} gave an action formulation of the membrane that starts by modifying the usual variational principle
\begin{equation}
    \delta_g S_{\text{total}}=\delta_g \left (S_{\text{in}} +S_{\text{out}} \right ),
\end{equation}
to 
\begin{equation}
    \delta_g S_{\text{total}}=\delta_g (S_{{\text{in}}}+ S_{{\text{surface}}})+\delta_g(S_{\text{out}}-S_{\text{surface}}),
\end{equation}
where the surface term refers to the black hole boundary that now is represented effectively by the membrane. The first and second parts of the action variations are assumed to be zero individually \cite{MembraneHorizonsBlackHolesNewClothes}. 

\subsubsection{The gravitational membrane}
Now, we rigorously find the variation on the stretched horizon.
\begin{equation}
    S_{\text{out}}=\frac{1}{16\pi}\int_\mathcal{M} d^4x\sqrt{-g} \,R+\frac{1}{8\pi}\oint_\mathcal{\partial M} d^3x\sqrt{\pm h}\,K,
\end{equation}
where we assumed $G_N=1$, $c=1$ and the second term is the Gibbons-Hawking boundary term. Using the Palatini's identity, one has \cite{BoundaryTermsoftheEinstein--HilbertAction}
\begin{equation}
    g^{\mu\nu}\delta R_{\mu\nu}=g^{\mu\gamma}\nabla_\gamma\big(g^{\lambda\nu}\nabla_\lambda\delta g_{\mu\nu}\big)-g^{\alpha\beta}\nabla_\mu\delta g_{\alpha\beta},
\end{equation}
where $\delta g_{\mu\nu}$ can be raised and lowered as an ordinary tensor. So:
\begin{gather}
    \int_\mathcal{M} d^4x\sqrt{-g}\, g^{\mu\nu}\delta R_{\mu\nu}
    =\int_\mathcal{\partial M} d^3x\sqrt{-h}\,n^\mu \,h^{\nu\alpha}\left(\nabla_\alpha\delta g_{\mu\nu}-\nabla_\mu\delta g_{\nu\alpha}\right)\equiv I.
\end{gather}
We choose the normal unit vector $n^\mu$ as outward pointing. Applying the Leibniz rule to the integrand gives
\begin{equation}
I=\int_\mathcal{\partial M} d^3x\sqrt{-h}\, h^{\mu\nu}\left[\left(\delta g_{\mu\alpha}\nabla_\nu n^{\alpha}-\delta g_{\mu\nu}\nabla_\alpha n^\alpha\right)+\left(\nabla_\alpha(n^\alpha\delta g_{\mu\nu})-\nabla_\nu(n^\alpha\delta g_{\alpha\mu})\right)\right].
\end{equation}

Using the definition of the extrinsic curvature as given in the paragraph above \eqref{extrinsiccurvaturedefinition}, one finds the variation of the action as the Brown-York quasilocal stress tensor.
\begin{equation}
    \delta S=\frac{1}{16\pi} \int_\mathcal{\partial M} d^3x\sqrt{-h}\big(Kh_{\mu\nu}-K_{\mu\nu}\big)\delta g^{\mu\nu}.
    \label{stretchedhorizontensor}
\end{equation}
Then, one has $t^{\text{stretched}}_{\mu\nu}=\frac{1}{8\pi}\left(Kh_{\mu\nu}-K_{\mu\nu}\right)\in T_p(\mathcal{H}_s)\otimes T_p(\mathcal{H}_s)$. On the stretched horizon $g^{\mu\nu}\vert_{\mathcal{H}_s}=h^{\mu\nu}$.To cancel the above nonzero boundary variation term, one must add the following:

\begin{equation}
    \delta S_{\text{surface}}=-\frac{1}{2}\int d^3 x\sqrt{-h}\,t^{\text{stretched}}_{\mu\nu}\delta h^{\mu\nu}.
\end{equation}
As it happens in electrodynamics where a surface charge induces a discontinuity in the field strength on the surface, $t^{\text{stretched}}_{\mu\nu}$ induces a discontinuity in the stretched horizon's extrinsic curvature $K_{\mu\nu}$ \cite{Parikh_original}. This discontinuity creates a junction on the surface that can be identified as the Israel junction condition \cite{Israel_junction_condition}:
\begin{equation}
    t^{\text{stretched}}_{\mu\nu}=\frac{1}{8\pi}\left([K]h_{\mu\nu}-[K]_{\mu\nu}\right),
\end{equation}
where $[K]=K^+-K^-$ such that $[K]$ is the difference between the external Universe embedding of $\mathcal{H}_s$. We should identify $K^-=0$ so that the stretched horizon interior to the black hole side is a flat embedding.
After all these considerations, one finds that $t_{\text{stretched}}^{\mu\nu}$ is not covariantly conserved: There is a source term and the equation reads as
\begin{equation}
    D_\nu t_{\text{stretched}}^{\mu\nu}=-{h}{^\mu}_{\lambda}T^{\lambda\gamma}n_\gamma.
\end{equation}
Hence the gravitational membrane acts like a fluid obeying Damour-Navier-Stokes equations on the spacelike cross section of the stretched horizon \cite{MembraneView}. One can write ${K}{^\mu}_{\nu}$ in terms of the surface gravity $\kappa$ and extrinsic curvature ${k}{^A}_{B}$ of the spacelike section of $\mathcal{H}_s$. To this end, one has $\nabla_\ell \ell=\kappa_r \ell$ where $\kappa_r$ is the normalized surface gravity at the horizon, which is related to the surface gravity $\kappa$ as $\kappa_r=N \kappa$ \cite{Parikh_original}.

Let $K_{AB}$ be the extrinsic curvature of the 2-space-like section of $\mathcal{H}_s$. Then, it can be separated into trace and traceless parts as
\begin{equation}
    K_{AB}=\sigma_{AB}+\frac{1}{2}\gamma_{AB}\Theta,
    \label{2.132}
\end{equation}
where $\sigma_{AB}$ is the shear tensor. Then finally, the stretched horizon stress tensor becomes \cite{MembraneHorizonsBlackHolesNewClothes}
\begin{flalign}
t^{AB}_{\text{stretched}}&=\frac{1}{8\pi}\left(-\sigma^{AB}+\gamma^{AB}(\frac{1}{2}\Theta+\kappa)\right),
\end{flalign}
which is the main formula that we shall use in what follows.

\section{ Membranes for the Modified Static and Stationary Metrics }

\subsection{A membrane for the Schwarzschild-type geometry}

Consider a spherically symmetric metric:
\begin{equation}
    ds^2=-fdt^2+f^{-1}dr^2+r^2d\Omega_2,
\end{equation}
where $f=1-\frac{2 m(r)}{r}$. We can the metric as\footnote{This construction follows \cite{Arslaniev_original}.}
\begin{equation}
    ds^2=-u_\mu u_\nu dx^\mu dx^\nu+n_\mu n_\nu dx^\mu dx^\nu+\gamma_{\mu\nu}dx^\mu dx^\nu,
\end{equation}
 with $u_\mu$ and $n_\mu$ chosen to satisfy $u_\mu u^\mu=-1$,  $n_\mu n^\mu=1$ while  $u_\mu n^\mu=0$ on the stretched horizon $\mathcal{H}_s$. At the event horizon $\mathcal{H}_{r\rightarrow r_H}$, the vectors $u$ and $n$ should be null. We know that $m(r)$ can only have global hairs hence it is of the form $m(r)=m-\frac{q^2}{2r}+\frac{\Lambda r^3}{6}$. For the sake of simplicity, let us consider the $\Lambda= q =0$ case. Then, the event horizon is located at $r_H=2 m$ for which $f(r_H)=0$.

On the 2D surface, we will use the coordinates $\{A,B\}=\{\theta,\phi\}$ hence $\gamma_{AB}=\text{diag}(r^2,r^2\sin^2\theta)$. 
The extrinsic curvature tensor for this geometry reads as 
\begin{equation}
K_{\mu\nu}=-\frac{1}{2}\frac{\partial_r f}{\sqrt{f}} u_\mu u_\nu+\frac{\sqrt{f}}{r}\gamma_{\mu\nu},
\end{equation}
of which the trace is $K=\frac{1}{2}\frac{\partial_r f}{\sqrt{f}}+2\frac{\sqrt{f}}{r}$.
In the $(2+1+1)$ splitting of the spacetime, the extrinsic curvature on $\mathcal{H}_s$ can be identified by choosing the lapse function $N=\sqrt{f}$ as a renormalization factor:
\begin{equation}
    K_{\mu\nu}\longrightarrow N^{-1}(k_{\mu\nu}+\kappa u_\mu u_\nu),
    \label{extrinsiccurvature}
\end{equation}
where $k_{\mu\nu}=\gamma_{\mu A}\gamma_{\nu B}k^{AB}$ is the extrinsic curvature of the 2D surface and $\kappa$ is the surface gravity \cite{Arslaniev_original}.

As $N\rightarrow  0$, the extrinsic curvature of the stretched horizon converges to the event horizon and $K_{\mu\nu}$ becomes proportional to the surface gravity. The trace of $K_{\mu\nu}$ diverges since $f$ has a pole at $r=r_H$.
\begin{flalign}
    &\lim_{N\to 0} K=\left.\frac{1}{2}\frac{\partial_r f}{\sqrt{f}} 
  \right\vert_{r_H}\rightarrow
    \text{Tr}(N^{-1}k_{\mu\nu}-N^{-1}\kappa u_\mu u_\nu)\vert_{(r=r_H)},\label{limitscalarcurvature}\\
    &\lim_{N\to 0} K_{tt}=-\left.\frac{1}{2}\frac{\partial_r f}{\sqrt{f}} \right\vert_{r_H}\rightarrow N^{-1}\kappa\vert_{r_H}\label{limitKtt}
\end{flalign}
Equations \eqref{extrinsiccurvature}\textendash\eqref{limitKtt} can be combined to find the stress tensor in terms of given parameters:
\begin{align}
    t^{\text{stretched}}_{\mu\nu}&=\frac{1}{8\pi N}\left((\Theta+\kappa)(\gamma_{\mu\nu}-u_\mu u_\nu)+\kappa u_\mu u_\nu- (\sigma_{\mu\nu}+\frac{1}{2}\Theta\gamma_{\mu\nu}) \right)&\notag\\
    &=\frac{1}{8\pi N}\left((\frac{1}{2}\Theta+\kappa)\gamma_{\mu\nu}-\Theta u_\mu u_\nu-\sigma_{\mu\nu}\right).
    \label{stretchedhorizonstresstensor}
\end{align}
One can compare this stretched stress tensor with a viscous fluid stress tensor
\begin{align}
    t^{\text{viscous}}_{\mu\nu}&=N^{-1}\rho u_\mu u_\nu+N^{-1}\gamma_{\mu A}\gamma_{\nu B}\big(P\gamma^{AB}-2\eta\sigma^{AB}-\zeta\Theta\gamma^{AB}\big)\label{viscoustensor}&\\&+\pi^A(\gamma_{\mu A}u_\nu+\gamma_{\nu B}u_\nu)\notag,&
\end{align}
with energy density $\rho$, pressure $P$, null geodesic expansion coefficient $\Theta$, bulk viscosity $\zeta$,  shear viscosity $\eta$, momentum density $\pi^A$, shear tensor $\sigma^{AB}$ near the event horizon. If we  identify \eqref{stretchedhorizonstresstensor}  and \eqref{viscoustensor} we get the following:
\begin{equation}
    \rho=-\frac{1}{8\pi}\Theta ,  \quad\eta=\frac{1}{16\pi},
    \quad P=\frac{\kappa}{8\pi} ,\quad \zeta=-\frac{1}{16\pi},\quad
    \pi^A=0.
    \label{transportcoefficientsstatic}
\end{equation}
Since we also have 
\begin{equation}
    t^{\text{stretched}}_{\mu\nu}=\frac{1}{8\pi}\left(\left(\frac{1}{2}\frac{\partial_r f}{\sqrt{f}}+\frac{\sqrt{f}}{r}\right)\gamma_{\mu\nu}-\frac{2\sqrt{f}}{r}u_\mu u_\nu\right),
\end{equation}
we have 
\begin{align}
\label{viscousstatic}
    \Theta=\frac{2}{r}f ,&& \sigma_{AB}=0,&&
    \kappa=\frac{\partial_r f}{2}&.
\end{align}
In particular, for the Schwarzschild geometry with $f=(1-\frac{2 m}{r})$, we have
\begin{equation}
    \Theta\vert_{r=r_H}=0 ,\quad \sigma_{AB}\vert_{r=r_H}=0,\quad
     \kappa\vert_{r=r_H}=\frac{1}{4 m}.
\end{equation}
We aim to generalize the static geometry as much as possible, if we restate the transport coefficients of generic static black holes, we can classify them by choosing the metric function $f(r)$. Observe that the surface gravity $\kappa$, energy density $\rho$, pressure $P$, and null expansion $\Theta$ will change for different choices of $f$. However, $\eta$, $\sigma^{AB}$, $\zeta$ will be a classification for spherical horizons and will be intact
for spherical horizons; in particular, the value of the bulk viscosity is negative showing that we are dealing with an unstable fluid.

\subsection{Doubly modified Schwarzschild metric}
Johannsen-Psaltis black hole is a parametrically deviated rotating black hole; before we work out its membrane construction, we would like to give the nonrotating version. This is because membrane paradigm analysis in the nonrotating limit ensures an easier detection of the transport coefficients, especially pressure according to truncation to the correct static limit. By taking the zero angular momentum limit to the JP black hole at hand we should be able to fix $\zeta=-\frac{1}{16\pi}$ condition \cite{Arslaniev_original}. This also implies that the nonrotating limit of the Johannsen-Psaltis black hole should have the same surface gravity as its doubly modified Schwarzschild counterpart. Now, we will present the doubly modified Schwarzschild metric, we show that equality at the level of surface gravity brings another constraint on the deformation function and relates $\epsilon_3 \simeq \alpha$. First, let us introduce the metric and find its transport coefficients:

\begin{equation}
    ds^2=-F \,h\,dt^2+\frac{h}{F\,g}dr^2+r^2\,h\,(d\theta^2+\sin\theta^2 d\phi),
\end{equation}
where
\begin{align*}
   F(r)=\frac{f(r)}{g(r)},&&
   h(r)=1+\frac{m^3   \epsilon _3}{r^3},&&
   f(r)=1-\frac{2 m}{r},&&
   g(r)=1+\frac{m^3  \alpha}{r^3},&&
\end{align*}
where we have a Schwarzschild-like causal structure with two seemingly distinct additional hairs, i.e., $\epsilon _3$ and $\alpha$.  In the older version of the JP metric \cite{A_metric_testing_johannsen}, one can see that the corrections to the Kerr metric were realized by applying the Newman-Janis algorithm to the modified Schwarzschild metric. Through this algorithm, the new hair is also complexified and naturally adapted to the Kerr-like metric. However, the metric in \cite{Johannsen_2013}, the Johannsen metric, is already in Klein-Gordon separable form and the deformations only depend on radial coordinate $r$. This gives one the freedom to choose those new deformations in such a way that the metric has the same event horizon radius as the Kerr black hole. Hence, currently, its static counterpart has not been given in the literature. To be able to fix the bulk viscosity as $\zeta=-\frac{1}{16 \pi}$, we choose its nonrotating limit to have the Schwarzschild radius.

One can realize that this metric has a similar causal structure as the Schwarzschild metric with the horizon at $r_H= 2 m$ as in the case of its rotating part has $r_H=m^2+\sqrt{m^2-a^2}$. Now, we can turn the paradigm's machinery on and find its dual fluid correspondence. Firstly, one can directly choose its spacelike normal vector as $n_\mu=\left\{0,\sqrt{\frac{h}{g F}},0,0\right\}$.
Then the Parikh-Wilczek type decomposition reads as
\begin{equation}
     u_\mu dx^\mu=\sqrt{F h}dt,\quad\quad
      n_\mu dx^\mu=\sqrt{\frac{h}{g F}}dr,
\end{equation}
and the 2D cross section of the black hole metric is 
\begin{equation}
    \gamma_{\mu\nu}dx^\mu dx^\nu=h\,r^2d\Omega^2_2.
\end{equation}
 The acceleration $a_\nu=n^\mu\nabla_\mu n_\nu$ vanishes
\begin{align}
    a_\nu&=n^\mu\left(\partial_\mu n_\nu-\Gamma^\gamma_{\mu\nu}n_\gamma\right)\notag=
    \sqrt{\frac{h}{g F}}{\delta}{^\mu}_{r}\left(\partial_\mu\left(\sqrt{\frac{h}{g F}}{\delta}{^r}_{\nu}\right)-\Gamma^\gamma_{\mu\nu}\left(\sqrt{\frac{h}{g F}}{\delta}{^r}_{\gamma}\right)\right)=0.&
\end{align}
This means momentum and the shear on the horizon are trivial which is expected for a static spacetime. 

\begin{align}
    &K_{tt}=-\frac{h\partial_r F+F \partial_r h}{2 \sqrt{\frac{h}{g F}}},
    \quad K_{rr}=0,\notag\\
    &K_{\theta\theta}=\frac{r \left(r \partial_r h+2 h\right)}{2 \sqrt{\frac{h}{g F}}},\quad
    \quad K_{\phi\phi}=\frac{r  \sin ^2 \theta   \left(r \partial_r h+2 h\right)}{2 \sqrt{\frac{h}{g F}}}.
 \end{align}
which reads as
\begin{equation}
K_{\mu\nu}=\sqrt{\frac{g}{F h}}\left(\left(\frac{ F\,  \partial_r h}{2 h}+\frac{ F }{r}\right)\gamma_{\mu\nu}-\left(\frac{1}{2}   \partial_r F-\frac{ F  \partial_r h}{2 h}\right)u_\mu u_\nu\right),
\end{equation}
of which the trace is
\begin{equation}
    K=\sqrt{\frac{g h}{ F}}\frac{\left(r h\partial_r F+F \left(3 r \partial_r h+4 h\right)\right)}{2 r h^2}.
\end{equation}
Then, the stretched horizon stress tensor \eqref{stretchedhorizonstresstensor}  with $\sigma_{\mu\nu}=0$ is 
\begin{equation}
     t_{\mu\nu}^{\text{stretched}}=\frac{1}{8\pi} \left(\frac{F g}{h}\right)^{\frac{1}{2}}\bigg(-  \left(\frac{\partial_r h}{h}+\frac{2}{r}\right) u_\mu u_\nu\notag+\left(\frac{1}{2}  \left(\frac{\partial_r h}{h}+\frac{2}{r}\right)+  \frac{\partial_r h} {2 h}+\frac{1}{2}\frac{ \partial_r F}{F}\right)\gamma_{\mu\nu}\bigg).
     \label{3}
\end{equation}
Note that in the limit of no deformation, the stretched horizon stress tensor reduces to Schwarzschild's membrane stress tensor given as 
\begin{equation}
    t_{\mu\nu}^{\text{Schwarzschild}}=\frac{1}{8\pi (F)^{\frac{1}{2}}}\left(-  \frac{2 F}{r} u_\mu u_\nu+\left( \frac{ F}{r}+ \frac{1}{2}  \partial_r F\right)\gamma_{\mu\nu}\right).
     \label{hgoes00}
\end{equation}
By using the transport coefficients given in \eqref{transportcoefficientsstatic}, we can  
proceed with the same analysis while choosing the nullness condition as
\begin{equation}
    \frac{1}{N}=\sqrt{\frac{1}{F h}},
\end{equation}
while the expansion, the shear tensor, and the surface gravity become
\begin{align}
    \Theta= \sqrt{g} F \left(\frac{\partial_r h}{h}+\frac{2}{r}\right),& &\sigma_{AB}=0,&&
     \kappa=\sqrt{g} F \frac{\partial_r h} {2 h}+\frac{1}{2} \sqrt{g} \partial_r F.
\end{align}
The stretched horizon is affected by deformations. The surface gravity as a function of $r$ becomes 
\begin{equation}
    \kappa=\frac{\alpha  m^3 r^3 (3 r-4 m)+m^3 \epsilon _3 \left(2 \alpha  m^4+8 m r^3-3 r^4\right)+2 m r^6}{2 r^2 \left(\alpha  m^3+r^3\right) \sqrt{\frac{\alpha  m^3}{r^3}+1} \left(m^3 \epsilon _3+r^3\right)},
  \end{equation}
  which at the event horizon reads as
  \begin{equation}
    \kappa_{r=2m}=\frac{1}{4m \sqrt{\frac{\alpha }{8}+1} },
    \label{surfacegravitydoublymodifid}
    \end{equation}
    restricting the deformation parameter $\alpha$ as $\alpha>-8$. On the other hand, the null expansion coefficient
    \begin{equation}
    \Theta_{r=2m}=\frac{(r-2 m) \left(m^3 \epsilon _3+4 r^3\right)}{2 r^2 \sqrt{\frac{\alpha  m^3}{r^3}+1} \left(m^3 \epsilon _3+r^3\right)}=0
\end{equation}
vanishes at the horizon as expected.
\subsection{The latest version of Johannsen-Psaltis spacetime}
The black hole introduced in \cite{A_metric_testing_johannsen} suffers from a violation of the strong rigidity theorem, and carries a chaotic geodesic equation \cite{chaotic2}. Its event horizon equation is given by a quintic equation that is not solvable by radicals. In \cite{A_metric_testing_johannsen}, instead of the event horizon, the authors studied the Killing horizon which is not the same thing when strong rigidity is violated. These problems make it hard for us to study its transport coefficients in the limit of the true horizon. In \cite{HeumannPsaltis}, the authors reconsidered the metric given in \cite{Johannsen_2013} and made some judicious choices of the metric functions. The new metric in \cite{HeumannPsaltis} violates the strong rigidity; however, it obeys the rotosurface theorem and the weak rigidity theorem that allow one to analyze the black hole angular momentum and surface gravity while equating the event horizon to the Killing horizon. In the latest work \cite{FeryalOzel}, the authors considered the form suggested in \cite{Johannsen_2013}. This will be the metric we shall study here given in local coordinates as
\begin{equation}
    ds^2=-\frac{S\mathcal{B}}{\mathcal{F}}dt^2-2a\frac{\Tilde{\Sigma}\mathcal{C}}{\mathcal{F}}\sin^2\theta dtd\phi
    +\frac{\Tilde{\Sigma}}{\Delta Z}dr^2\\+\Tilde{\Sigma} d\theta^2
    +\frac{\Tilde{\Sigma}\mathcal{D}\sin^2\theta}{\mathcal{F}}d\phi^2,
\end{equation}
where
\begin{align}
    \mathcal{B}&=\Delta-a^2 B^2\sin^2\theta,\quad
    \mathcal{C}=(r^2+a^2)A B-\Delta,\notag\\
    \mathcal{D}&=(r^2+a^2)^2 A^2-a^2\Delta\sin^2\theta,\quad
    \mathcal{F}=\left((r^2+a^2)A-B\sin^2\theta\right)^2.
 \end{align}
 While choosing the metric functions as
\begin{align}
    &F_t=\frac{\sqrt{S \left(\Delta-a^2 B^2 \sin ^2 \theta  \right)}}{\left(\left(a^2+r^2\right) A-a^2 B \sin ^2 \theta  \right)} ,
    \quad F_\phi=\frac{\sqrt{\sin ^2 \theta   S \left(\left(a^2+r^2\right)^2 A^2-a^2 \sin ^2 \theta   \Delta\right)}}{\left(\left(a^2+r^2\right) A-a^2 B \sin ^2 \theta  \right)},\notag\\
    &F_r=\sqrt{\frac{S}{\Delta Z}},\quad\quad\quad
    \omega=\frac{a \sin ^2 \theta   S \left(\Delta-\left(a^2+r^2\right) A B\right)}{\left(\left(a^2+r^2\right) A-a^2 B \sin ^2 \theta  \right) \sqrt{S \left(\Delta-a^2 B^2 \sin ^2 \theta  \right)}},&
\end{align}
and functions
\begin{align*}
     \Sigma=r^2+a^2 \cos^2\theta,
    &&S(r,\theta)= \Sigma+\sum_{n=3}^{\infty} \epsilon_n\left(\frac{m^{n}}{r^{n-2}}\right),
    &&\Delta=r^2+a^2-2 m r,\\
    A=1+\sum_{n=3}^{\infty} \alpha_n\left(\frac{m}{r}\right)^n,
   && B=1+\sum_{n=2}^{\infty} b_n\left(\frac{m}{r}\right)^n,
    &&Z=1+\sum_{n=2}^{\infty} z_n\left(\frac{m}{r}\right)^n,
\end{align*}
 the metric deformations are kept dimensionless and for weak field tests, one needs to set $\epsilon_2=\alpha_2=0$. This form of the metric is asymptotically flat, reproduces the Newtonian effects in the limit and satisfies the PPN constraints. We will consider only the lowest-order deformation of the metric, not the whole summation of the deformation parameters. One can clearly understand that the main difference between this metric and the metric in \cite{A_metric_testing_johannsen} is that it has more hair introduced via deformations; however, it still has the Kerr event radius, that is, $g^{rr}\vert_{r_H}=0$ with $r_H=m+\sqrt{m^2+a^2}$. In our analysis, we should consider three different aspects:
\begin{enumerate}
    \item We will fix the bulk viscosity of the rotating solution to be the same as the nonrotating one, i.e., $\zeta=-\frac{1}{16\pi}$.  
    \item All the transport coefficients of the modified black hole should reduce to those of the Kerr black hole in the no deformation limit.
    \item We should check the finiteness of the transport coefficients for all deformation parameters.
\end{enumerate}
 Using the metric functions, the generic metric can be recast as
\begin{equation}
    ds^2=-F_t^2dt^2-2\omega F_tdtd\phi+F^2_\phi d\phi^2+ F_r^2dr^2+\Sigma d\theta^2,
    \label{3.61}
\end{equation}
or it can be written as \cite{Kerrintroduction}
\begin{equation}
    ds^2=-(F_tdt^2+\omega d\phi)^2+ F_r^2dr^2+\Sigma d\theta^2+(F_\phi^2+\omega)d\phi^2.
     \label{266}
\end{equation}
Now, we should decompose this metric in the $(2+1+1)$ form:
\begin{equation}
    ds^2=\left(-u_\mu u_\nu+n_\mu n_\nu+\gamma_{AB}{e}{^A}_{\mu}{e}{^B}_{\nu}\right)dx^\mu dx^\nu.
    \label{decomposedmetricform}
\end{equation}
Let $u_\mu dx^\mu=F_tdt+\omega d\phi$, $n_\mu dx^\mu=F_r dr$.
The structure of this metric can be put in the form of \eqref{3.61} such that one can directly start to calculate the factors relevant to the membrane description. {The results for the relevant expressions are cumbersome; therefore we delegate these to the Appendix; here, we only give the pressure in the relevant limits. } 
The pressure generically
is of the form $P(r,\theta;\alpha,z,b, a,\epsilon_3)$; and in the limit $z\rightarrow 0$ and $a\rightarrow 0$, it reduces to
\begin{equation}
    P_{|_{z\to 0,a\to 0}}=\frac{2 m r^4 \left(\alpha  m^2 (3 r-5 m)+r^3\right)+m^3  r \epsilon _3 \left(-4 \alpha  m^4+3 \alpha  m^3  r+8 m r^3-3 r^4\right)}{16 \pi  \left(\alpha  m^3 +r^3\right) \left(m^3  \epsilon _3+r^3\right){}^2},\notag\\
\end{equation}
 which at the Schwarzschild horizon, becomes 
\begin{equation}
     P_{|_{r \to 2 m}}=\frac{1}{4 m}\frac{1}{ \left(1+\frac{\epsilon _3}{8}\right)}.
\end{equation}
So, this is the pressure obtained in the nonrotating limit of the JP metric: it depends on the deformation parameter $\epsilon_3$. On the other hand, in \eqref{surfacegravitydoublymodifid} we obtained an apparently different value for the surface gravity when we directly did the calculation in the static black hole metric. Therefore, ostensibly the no-rotation case and the limit of no rotation seem to yield different results. This is not acceptable; hence, we should identify these two relations, which necessarily fix one of the deformation parameters in terms of the other as
\begin{align*}
&\kappa_{\text{nonrotating}} \overset{!}{=}\kappa_{\text{doubly modified}}\implies
    \frac{1}{4 m \left(\frac{\epsilon _3}{8}+1\right)}=\frac{1}{4 m}\sqrt{\frac{1}{\frac{\alpha }{8}+1}},
\end{align*}
\begin{equation}
    \epsilon _3=  8\left(\sqrt{1+\frac{\alpha}{8}}-1\right), \hskip 1 cm \alpha > -8.
    \label{epsilonalpha}
\end{equation}
This identification also allows us to fix the value of the bulk viscosity to be $\zeta=-\frac{1}{16\pi}$, in the static limit of the JP black hole, {while the shear viscosity is $\eta=\frac{1}{16\pi}$} and uniquely solve the transport coefficients for this case \cite{Arslaniev_original}. There are {three things to note here: First, the two additional hairs $\alpha$ and $\epsilon_3$ are not independent; Second, the effective membrane description is valid only for the region $\epsilon_3 > -8$. It is interesting to note that the critical point $\epsilon_3=-8$ was found before in a completely different context \cite{A_metric_testing_johannsen}. {Finally the ratio $\frac{\zeta}{\eta}=-1\leq\frac{2}{3}$ is consistent with the bound computed from the AdS/CFT correspondence \cite{Gubser:2008sz}.}

In Fig. \ref{figure1}, pressure versus the radial coordinate is plotted for all $r$ for the JP black hole. There is a discontinuity at the location of the ergosphere, and the pressure vanishes at the inner and outer horizons while diverging at the central singularity. It also asymptotically vanishes as $r\rightarrow\infty$. To compare we have also plotted as in Fig. \ref{figure2} the pressure of the fluid for the Kerr black hole (see \cite{Arslaniev_original} where this plot was first given). The JP black hole and the Kerr black hole have similar $P$ vs $r$ graphs: The main difference is the sign of the pressure at the central singularity. In Fig. \ref{figure3}, we plotted the $P$ vs $r$ graph for the doubly modified Schwarzschild black hole. It is similar to the ordinary Schwarzschild case.

\begin{figure}[ht]
\includegraphics[width=10cm]{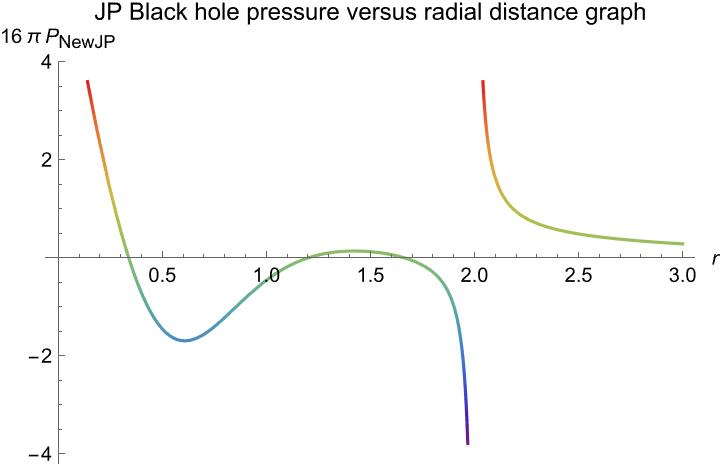}
\caption{{First a caution about this and the other figures; strictly speaking the membrane does not extend inside of the horizon. But it is remarkable to see that analytically continuing the pressure to the inside of the horizon, the fluid manages to capture the important surfaces of the black hole. For this reason, we plotted the pressure versus the radial coordinate over the whole range up to the central singularity.} This figure represents the scaled pressure as a function of the radial coordinate $r$ for the choices $\alpha=1 ,m=1, b=0, a=0.75, \epsilon _3\to 6 \sqrt{2}-8$. For these choices the Kerr horizon radii read as $r_{\text{H}_{outer}}= 1.66$ while $r_{\text{H}_{inner}}= 0.33$. One can see that at the ergosphere radius at the equator, $r_{\text{ergosphere}}=2$, there is a discontinuity in the pressure. One observes that the pressure diverges positively at the central singularity, discontinuous at the ergosphere radius at the equatorial plane, and asymptotically zero at infinity while vanishing at both the inner and outer event horizons. }
\label{figure1}
\end{figure}

\begin{figure}[ht]
\includegraphics[width=10cm]{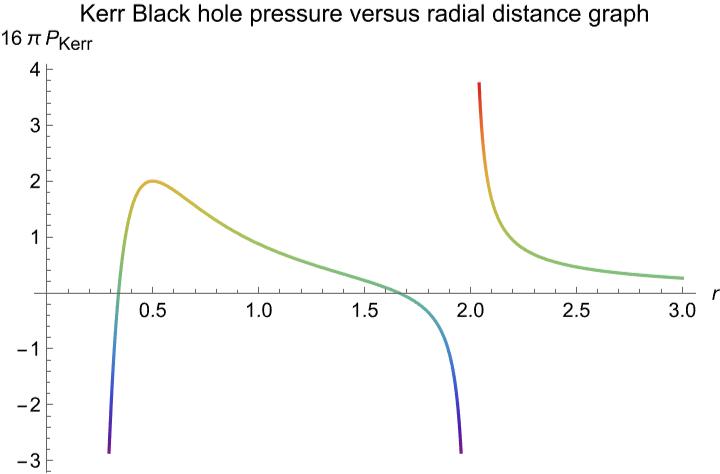}
\caption{ This figure represents the Kerr black hole's dual fluid pressure versus radial coordinate $r$, for the choice $m=1, a=0.75$, the Kerr radius $r_{\text{H}_{outer}}= 1.66$ while $r_{\text{H}_{inner}}= 0.33$. One can see that at the ergosphere radius on the equator, there is a discontinuity at $r_{\text{ergosphere}}=2$.  The pressure diverges negatively at the central singularity, discontinuous at the ergosphere radius at the equatorial plane, and is asymptotically zero at infinity while vanishing at both the inner and outer event horizons. }
\label{figure2}
\end{figure}

\begin{figure}[ht]
\includegraphics[width=10cm]{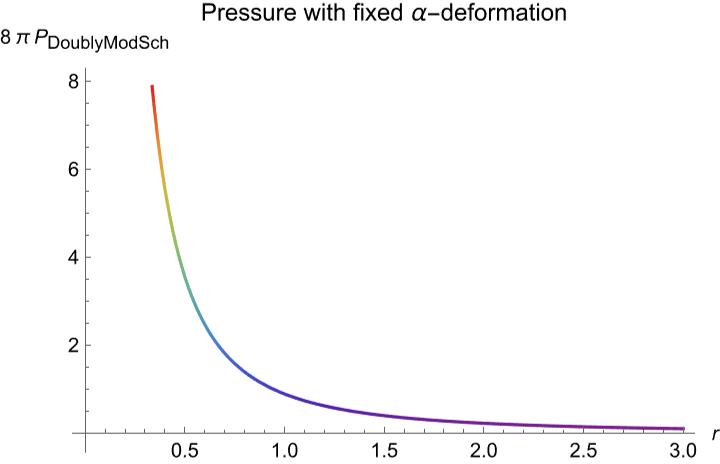}
\caption{This figure represents pressure values of the doubly modified Schwarzschild black hole scaled with $8\pi$ when we choose $m=1, a=0,\alpha=1$, the Schwarzschild radius at $r_{\text{H}}= 2$.  As we expected, the function is monotonically decreasing as $r$ gets larger reaching zero at asymptotic infinity, while positively diverging at the central singularity. }
\label{figure3}
\end{figure}
 Now, let us check the transport coefficient that carries the energy flux into the null surface, i.e., the null expansion $\Theta$. When the metric coefficients are plugged in one has:
\begin{flalign}
    \Theta&=\frac{\left(\frac{m^2 z}{r^2}+1\right)}{I(r,\theta)} \bigg(\frac{2 a^2 \left(b m^2+r^2\right) \left(2 a^2 b m^2+b m^2 r (3 r-5 m)+r^3 (r-m)\right) \left(a^2 \cos ^2 \theta +r^2\right)}{a^2 r \left(b m^2+r^2\right)^2-r^5 \csc ^2 \theta   \left(a^2+r (r-2 m)\right)}\notag\\&+\frac{\left(a^2+r (r-2 m)\right) \left(2 r^3-m^3   \epsilon _3\right) \left(a^2 \cos ^2 \theta +r^2\right)}{r \left(a^2 r \cos ^2 \theta +m^3   \epsilon _3+r^3\right)}+2 r \left(a^2+r (r-2 m)\right)\bigg),&&
\end{flalign}
where $I(r,\theta)=2 \left(a^2 \cos ^2 \theta +r^2\right)^2$ and we have also included the $z$-deformation dependence.
On the equatorial plane, in the nonrotating limit one gets 
\begin{align}
    &\Theta_{\theta \to \frac{\pi }{2},a\to 0}=\frac{1}{2 r^4}\left(\frac{m^2 z}{r^2}+1\right) \left(\frac{r^2 (r-2 m) \left(2 r^3-m^3   \epsilon _3\right)}{m^3   \epsilon _3+r^3}+2 r^2 (r-2 m)\right),
\end{align}
where we now can insert \eqref{epsilonalpha} to the null expansion while setting the deformation parameter $z=0$:
\begin{align}
   \Theta_{\theta \to \frac{\pi }{2},a\to 0} &=\left(1-\frac{2 m}{r}\right) \left(\frac{r^3-\left(\frac{1}{\frac{1}{4} \sqrt{\frac{1}{\frac{\alpha }{8}+1}}}-4\right) m^3}{2 \left(\frac{1}{\frac{1}{4} \sqrt{\frac{1}{\frac{\alpha }{8}+1}}}-4\right) m^3+r^3}+1\right).
\end{align}
 We can check the vanishing deformation limit and limit at the critical deformation parameter,
\begin{align}
    &\Theta_{|_{\alpha=0}}=2\left(1-\frac{2 m}{r}\right),\label{nullexpSCH}\\
    &\Theta_{|_{\alpha=-8}}=\frac{2 \left(r^3-2 m^3\right)}{r \left(4 m^2+2 m r+r^2\right)}\label{nullexpatcriticality}.
\end{align}
Equation \eqref{nullexpSCH} is the exact behavior found for the Schwarzschild black hole \cite{Arslaniev_original}.  

\section{Conclusions and Discussions}
Using the Parikh-Wilczek \cite{MembraneHorizonsBlackHolesNewClothes}  action formulation, we developed a membrane description of the 
Johannsen-Psaltis black hole, which is a phenomenologically viable deformation of the Kerr metric that is amenable to test the Kerr hypothesis along with the no-hair theorem.  As alluded to in the Introduction, a membrane is an effective description of a black hole as it appears to an outside observer. As such, it replaces the teleological concept of the black hole event horizon  (a null hypersurface), which is not possible to probe by transient observers like us with a timelike fluid. The definition of the membrane is such that one might wonder if one is bartering the teleological event horizon with a tautological membrane that impersonates the black hole. This is not the case, but if it were the case, it would still not be so terrible because the membrane is local in time and has proved to provide an intuitive understanding of complicated phenomena such as the relativistic jet production by rotating black holes and their accretion disk. {In this work, we have not studied the jet-production mechanism through the membrane paradigm, but the fact that one can construct a membrane for the JP metric for some deformation parameters is useful. Moreover, seemingly independent deformation parameters must be related for the membrane paradigm to work. We have ruled out some deformation parameters of the JP metric via the membrane paradigm. We have also shown that the ratio of the bulk viscosity to shear viscosity remains within the bounds predicted by the AdS/CFT dictionary. }

As we have seen here the membrane can accommodate extra hairs yet constrain the values of these hairs. Our main task was to understand if a complicated metric such as the JP metric can fit into the membrane description, and we have seen that this is possible. If we analytically continue the pressure as a function of the radial coordinate $r$, we observe several interesting behaviors: At the location of the ergosphere on the equator, the pressure diverges just like the pressure of a Van der Waals gas diverges at the volume of the molecule. Therefore, the fluid is aware of the size of the ergoregion of the black hole. This might have interesting applications in astrophysical black holes.  For example, in \cite{BlandfordZnajek}, it was shown that the Blandford-Znajek \cite{Blanford-ZnajekOriginal} that explains the relativistic jet production of rotating black holes with accretion discs is related to the ergosphere of the black hole and not to the event horizon. If the jet production is to be explained by the membrane paradigm it is clear that our construction above lends support to the computations of \cite{BlandfordZnajek}.
\section{Acknowledgments}
We would like to thank Professor Dr. Sakir Erkoc for their support of this work.

\section{Appendix: Details of Calculations for the Construction of the Membrane}

The components of the extrinsic curvature of the metric \eqref{decomposedmetricform} are found to be
\begin{flalign}
    K_{tt}&=-\frac{-2 S}{W(r,\theta)} \big(\left(\Delta-a^2 B^2 \sin ^2 \theta  \right) \left(\left(a^2+r^2\right) \partial_r A\notag -a^2 \sin ^2 \theta   \partial_r B+2 r A\right)\notag\\&+\partial_r S \left(\left(a^2+r^2\right) A-a^2 B \sin ^2 \theta  \right) \left(\Delta-a^2 B^2 \sin ^2 \theta  \right)\notag\\&+S \left(\left(a^2+r^2\right) A-a^2 B \sin ^2 \theta  \right) \left(\partial_r \Delta-2 a^2 B \sin ^2 \theta   \partial_r B\right)\big),&&
\end{flalign}
\begin{flalign}
    K_{t\phi}&=K_{\phi t}=\frac{1}{W(r,\theta)}((a \sin ^2 \theta(A (S ((a^2+r^2) B (a^2 \sin ^2 \theta   \partial_r B-(a^2+r^2) \partial_r A )\notag\\&-2 a^2 r B^2 \sin ^2 \theta  -(a^2+r^2) \partial_r \Delta+4 r \Delta)-(a^2+r^2) \partial_r S (a^2 B^2 \sin ^2 \theta  +\Delta))\\&+2 \Delta S ((a^2+r^2) \partial_r A -a^2 \sin ^2 \theta   \partial_r B)+(a^2+r^2) A^2 (S ((a^2+r^2) \partial_r B-2 r B)\notag\\&+(a^2+r^2) B \partial_r S)-a^2 (a^2+r^2) B^2 \sin ^2 \theta   S \partial_r A +a^2 B \sin ^2 \theta   (\Delta \partial_r S+S \partial_r \Delta)))),\notag&&
    \end{flalign}
    \begin{flalign}
     K_{\theta\theta}&=\frac{\partial_r S}{2 \sqrt{\frac{S}{\Delta Z}}},\quad
       K_{\phi\phi}=\frac{1}{E(r,\theta)}\left(\sin ^2 \theta   S \left(\left(a^2+r^2\right)^2 A^2-a^2 \sin ^2 \theta   \Delta\right)\right),\quad
       K_{rr}=0,&&      
\end{flalign}

where we introduced the following functions:
\begin{flalign}
    W(r,\theta)&=2 \sqrt{\frac{S}{\Delta Z}} \left(\left(a^2+r^2\right) A-a^2 B \sin ^2 \theta  \right)^3,\notag\\
    E(r,\theta)&=2 \sqrt{\frac{S}{\Delta Z}} \left(\left(a^2+r^2\right) A-a^2 B \sin ^2 \theta  \right)^2.
\end{flalign}
The trace of the extrinsic curvature is given as
\begin{flalign}
 K&= \frac{1}{R(r,\theta)}\big(3 \Delta \partial_r S \left(\left(a^2+r^2\right) A-a^2 B \sin ^2 \theta  \right)+S \big(\partial_r \Delta \left(\left(a^2+r^2\right) A-a^2 B \sin ^2 \theta  \right)\notag\\&-2 \Delta \left(\left(a^2+r^2\right) \partial_r A -a^2 \sin ^2 \theta   \partial_r B+2 r A\right)\big)\big),&&
\end{flalign}
where
\begin{equation}
    R(r,\theta)=2 \Delta S \left(\left(a^2+r^2\right) A-a^2 B \sin ^2 \theta  \right) \sqrt{\frac{a^2 \cos ^2 \theta +r^2}{\Delta Z}}.
\end{equation}
Following the construction of a generic Kerr-like membrane paradigm algorithm, one can find the other transport coefficients. For instance, the nonzero components of the shear tensor turn out to be
\begin{align}
  \sigma_{\theta\theta}&=\frac{a^2 B Z \left(2 \Delta \partial_r B-B \partial_r \Delta\right)}{4 a^2 B^2-4 \csc ^2 \theta   \Delta},&
\sigma_{\phi\phi}&=-\frac{a^2 B \Delta Z \left(B \partial_r \Delta-2 \Delta \partial_r B\right)}{4 \left(a^2 B^2-\csc ^2 \theta   \Delta\right)^2},&
\end{align}
and the nonzero component of the momentum becomes
\begin{small}
\begin{align}
  &\pi^\phi=-\frac{a \left| \left(a^2+r^2\right) A-a^2 B \sin ^2 \theta \right|}{D(r,\theta)}  \bigg(-\left(a^2+r^2\right)^2 A^2 \bigg(B \Delta \partial_r Z \left(a^2 B^2 \sin ^2 \theta -\Delta\right)\notag\\&+Z \left(a^2 B^3 \sin ^2 \theta  \partial_r \Delta-2 \Delta^2 \partial_r B\right)\bigg)+A \bigg(\Delta \bigg(-2 a^2 \left(a^2+r^2\right) B \sin ^2 \theta  Z \partial_r B \left(a^2 B^2 \sin ^2 \theta +\Delta\right) \notag\\&+4 r Z \left(\Delta-a^2 B^2 \sin ^2 \theta \right)^2+\left(a^2+r^2\right) \partial_r Z \left(a^4 B^4 \sin ^4 \theta -\Delta^2\right)\bigg)\\&+\left(a^2+r^2\right) Z \partial_r \Delta \left(a^4 B^4 \sin ^4 \theta +3 a^2 B^2 \sin ^2 \theta  \Delta-2 \Delta^2\right)\bigg)\notag\\&+\Delta \bigg(-a^4 B^3 \sin ^4 \theta  \left(\Delta \partial_r Z+3 Z \partial_r \Delta\right)+4 a^2 B^2 \sin ^2 \theta  \Delta Z \bigg(a^2 \sin ^2 \theta  \partial_r B-\left(a^2+r^2\right) \partial_r A\bigg)\notag\\&+2 \Delta^2 Z \left(\left(a^2+r^2\right) \partial_r A-a^2 \sin ^2 \theta  \partial_r B\right)+a^2 B \sin ^2 \theta  \Delta \left(\Delta \partial_r Z+2 Z \partial_r \Delta\right)\notag\\&+2 a^4 \left(a^2+r^2\right) B^4 \sin ^4 \theta  Z \partial_r A\bigg)\bigg) ,\notag
\end{align}
\end{small}
where $D(r,\theta)=32 \pi  \Delta^2 Z \sqrt{\frac{S}{\Delta Z}} \left(\left(a^2+r^2\right) A-a^2 B \sin ^2 \theta \right)^3 \sqrt{S \left(\Delta-a^2 B^2 \sin ^2 \theta \right)}$.

The null-expansion is
\begin{align}
    &\Theta=\frac{Z}{2 S^2} \left(\frac{a^2 B S \left(B \partial_r \Delta-2 \Delta \partial_r B\right)}{a^2 B^2-\csc ^2 \theta   \Delta}+2 \Delta \partial_r S\right),
\end{align}
The pressure reads as follows:
\begin{align}
   P&= \frac{\Delta Z}{16 \pi  S^2} \bigg(-\frac{2 S \left(\left(a^2+r^2\right) \partial_r A -a^2 \sin ^2 \theta   \partial_r B+2 r A\right)}{\left(a^2+r^2\right) A-a^2 B \sin ^2 \theta  }\notag\\&+\frac{S \left(\partial_r \Delta-2 a^2 B \sin ^2 \theta   \partial_r B\right)}{\Delta-a^2 B^2 \sin ^2 \theta  }+\partial_r S\bigg).&&  
   \label{feryal_pressure}
\end{align}
 When the metric coefficients are plugged to \eqref{feryal_pressure} it becomes
\begin{flalign}
    P&=\frac{1}{R(r,\theta)}\bigg(\left(a^2+r (r-2 m)\right) \left(\frac{m^2 z}{r^2}+1\right) \bigg(\frac{1}{Y(r,\theta)}\bigg(\frac{4 a^2 b m^2 \sin ^2 \theta   \left(b m^2+r^2\right)}{r^5}\notag\\&-2 m+2 r\bigg) \times  \notag \left(a^2 \cos ^2 \theta +\frac{m^3   \epsilon _3}{r}+r^2\right)-\frac{2}{U(r,\theta)} \bigg(a^2 r \cos ^2 \theta +m^3   \epsilon _3+r^3\bigg)\\&\times \bigg(2 a^2 b m^2 r \sin ^2 \theta  -\alpha  m^3   \left(3 a^2+r^2\right)+2 r^5\bigg)-\frac{m^3   \epsilon _3}{r^2}+2 r\bigg)\bigg),&&
\end{flalign}
where 
\begin{align}
    &R(r,\theta)=16 \pi  \left(a^2 \cos ^2 \theta +\frac{m^3   \epsilon _3}{r}+r^2\right)^2,\notag\\
    &Y(r,\theta)=a^2 \sin ^2 \theta   \left(-\left(\frac{b m^2}{r^2}+1\right)^2\right)+a^2+r (r-2 m),\\
    &U(r,\theta)=r^2 \left(\left(a^2+r^2\right) \left(\alpha  m^3  +r^3\right)-a^2 r \sin ^2 \theta   \left(b m^2+r^2\right)\right).\notag
\end{align}

\end{document}